\documentclass[epj]{svjour}
\usepackage{epsf,epsfig,psfig,float,amssymb,stmaryrd,latexsym}
\usepackage{graphics,psfrag}
\newcommand{\vs}{\vspace{-0.0cm}}
\newcommand{\beq}{\begin{equation}}
\newcommand{\eeq}{\end{equation}}
\newcommand{\beqa}{\begin{eqnarray}}
\newcommand{\eeqa}{\end{eqnarray}}
\newcommand{\nn}{\nonumber \\ }
\newcommand{\bma}{\begin{array}{cc}}
\newcommand{\ema}{\end{array}}
\newcommand{\Mpz}{M_\pi^2}
\newcommand{\Mp}{M_\pi}

\def\3{{\ss}}

\def\vek #1 {\overrightarrow {#1}}
\newcommand{\fet}[1]{\mbox{\boldmath $#1$}}

\begin{document}
\title{Improving the convergence of the chiral expansion for nuclear forces I:  Peripheral 
phases}

\author{E. Epelbaum \inst{1} \thanks{email: 
                           evgeni.epelbaum@tp2.ruhr-uni-bochum.de}
\and
W. Gl\"ockle \inst{1} \thanks{email:
                           walter.gloeckle@tp2.ruhr-uni-bochum.de}
\and
Ulf-G. Mei{\ss}ner \inst{2} \thanks{email: 
                           meissner@itkp.uni-bonn.de}
}
\institute{ 
  Ruhr-Universit\"at Bochum, Institut f{\"u}r
  Theoretische Physik II, D-44870 Bochum, Germany
\and
Universit\"at Bonn,
Helmholtz--Institut f\"ur Strahlen-- und Kernphysik (Theorie), 
Nu{\ss}allee 14-16,
D-53115 Bonn 
}
\date{Received: date / Revised version: date}
%
\abstract{
We study the two-pion exchange potential at next-to-next-to-leading order in
chiral effective field theory. We propose a new cut--off scheme for the pion
loop integrals based on  spectral function regularization. We show that this
method allows for a consistent implementation of constraints from 
pion--nucleon scattering. It leads to a much improved description of 
the partial waves with
angular momentum $l \geq 2$ as compared to the calculation utilizing  
dimensional regularization.
}
\maketitle
\section{Introduction}
\def\theequation{\arabic{section}.\arabic{equation}}
\setcounter{equation}{0}

Effective field theory (EFT) has become a standard tool for analyzing the
chiral structure of Quantum Chromodynamics (QCD) at low energies, 
where the perturbative expansion in powers  
of the coupling constant cannot be used. The chiral symmetry of QCD is
spontaneously broken and the corresponding Goldstone bosons can be 
identified with the pions, if one considers the two flavor sector of the
up and down quarks as done here. 
It is a general property of Goldstone bosons 
that their interactions become weak for small momenta. Chiral Perturbation 
Theory (CHPT) is the effective field theory of the Standard Model  which allows 
to describe the interactions of pions and between pions and matter
fields (nucleons, $\rho$--mesons, $\Delta$--resonances, 
$\ldots$) in a systematic way. This is achieved via an expansion of
scattering amplitudes and transition currents in powers of a
 low--momentum scale $Q$ associated 
with small external momenta and with the pion (light quark) mass. 
Here, small means with respect to the scale of chiral symmetry breaking of the 
order of 1~GeV. Pion loops are naturally incorporated and all corresponding 
ultraviolet divergences can be absorbed at each fixed order in the chiral 
expansion by counter terms of the most general chiral invariant
Lagrangian.  

This perturbative scheme works well in the pion and
pion--nucleon sector, where the interaction vanishes  
for vanishing external momenta in the chiral limit, for some
early reviews see e.g. \cite{Ulf,Pich,BKMrev,Ecker}.
The situation in
the few--nucleon sector is much more complicated. 
The main difficulty in the direct application 
of the standard methods of  CHPT to the two--nucleon (2N) system
is due to the non-perturbative aspects of the problem, the unnaturally large S-wave
scattering lengths and the shallow nuclear bound states.
One possible way to deal with this difficulty has been suggested by Weinberg, who  
proposed to apply CHPT to the kernel of the corresponding integral
equation for the scattering amplitude, 
which can be viewed as an effective nucleon--nucleon (NN) potential 
\cite{Weinb1,Weinb2}.

The first quantitative realization of the above idea has been
carried out by  Ord\'o\~nez and co--workers, 
who derived an (energy-dependent) NN potential and performed a 
numerical analysis of the 2N system \cite{Ordonez96}. The 
energy-independent representation of the chiral NN potential, which 
can be applied much easier in few--nucleon 
calculations, has been derived in \cite{Friar94,Kaiser97,EGM98}.
At leading order (LO) in the chiral expansion the potential 
is given by the well established one--pion exchange (OPE) and 
two contact forces without derivatives. 
At next--to--leading order (NLO) OPE gets renormalized and 
the leading two--pion exchange (TPE) diagrams as well as seven 
more contact operators appear. At NNLO, one has to include
subleading TPE with one insertion of dimension two pion--nucleon 
vertices (the corresponding low--energy constants (LECs)  are 
denoted by $c_{1,3,4}$, we adhere to the notation of Ref.~\cite{BKMrev}). 
Notice that no new contact forces contribute at this order. 
While the  pion exchanges are governed by the underlying chiral
symmetry, the contact forces represent our ignorance
of the short--range physics and are not restricted by chiral symmetry. 

The LECs $c_{1,3,4}$ enter the expressions for the pion--nucleon 
($\pi$N) scattering amplitude at subleading ($Q^2$) and higher orders 
and thus can be determined from the $\pi$N data. 
From the $Q^2$ analysis \cite{BKM95} one gets:
$c_1 = -0.64\,,  c_3 = -3.90\,, c_4=2.25\;$ 
(here and in what follows the values of the $c_i$'s are given in 
GeV$^{-1}$). The values obtained from various $Q^3$ analyses 
\cite{BKM95,BKM97,Moi98,FMS99,Paul} are in the ranges:
$c_1 =-0.81 \ldots -1.53  \,,  c_3 = -4.70 \ldots -6.19 \,, c_4=
3.25 \ldots 4.12\;$.
These bands are also consistent with expectations from resonance 
saturation \cite{BKM97}. Notice that the  numerical values of
$c_3$ and $c_4$ at both orders $Q^2$ and $Q^3$  are quite large, which can 
be partially explained by the fact that the LECs $c_{3,4}$ are to a 
large extent saturated by the $\Delta$--excitation.
This implies that a new and smaller scale, namely $m_\Delta - m \sim 293$ 
MeV, enters the values of these constants in EFT without explicit $\Delta$, 
see \cite{BKM97}. At fourth order, these LECs get modified by quark mass
dependent contributions \cite{FM4} and also are affected when electromagnetic
corrections are included \cite{FM3}. These modifications go beyond the accuracy
of the calculations performed here but should be kept in  mind if one wants to go
to higher orders and/or systematically includes isospin violation.

The large numerical values of the $c_i$'s lead to dramatic consequences 
in few--nucleon systems \cite{Epe02}. The resulting subleading TPE
correction calculated using dimensional regularization \cite{Kaiser97} (or 
equivalent schemes) turns out to be very strong already at intermediate 
distances $r \sim 1-2$ fm. This could, in principle, be compensated by 
the corresponding contact terms. However, such a compensation at NNLO is 
only possible in low partial waves (i.e. in S--, P--waves as well as for 
$\epsilon_1$) since the contact terms do not contribute to D-- and higher 
partial waves at this order. The D-- and F--waves may therefore serve as 
a sensitive test of chiral TPE exchange, as suggested by 
Kaiser et al. in \cite{Kaiser97}, since higher partial waves are strongly 
dominated by OPE and less sensitive to TPE (as it is known since long, for
an early nonperturbative approach see \cite{GGIP}).
The conventional scenario of nuclear forces represented by existing OBE 
models and various phenomenological potentials suggests that the  D-- and 
higher partial wave NN interactions are weak enough to be treated 
perturbatively, see \cite{Epe02}. Clearly, in such a framework one can not
describe the low partial waves that show strong nonperturbative effects.
Under this assumption
Kaiser et al.~\cite{Kaiser97} applied 
chiral EFT to perform a parameter--free calculation for the  neutron--proton 
({\it np}) D-- and higher partial waves and found rather poor convergence 
of the chiral expansion already at surprisingly low energies.
While the LO and NLO predictions, which correspond to OPE and 
to OPE accompanied by the leading TPE corrections, already agree reasonably well with 
the Nijmegen phase shift analysis \cite{Stoks93} (NPSA), taking into account 
subleading TPE at NNLO governed by the LECs $c_{1,3,4}$  destroys that 
agreement and leads to 
strong deviations from the data for $E_{\rm lab} > 50$ MeV in D--waves and for 
$E_{\rm lab} > 150$ MeV in F--waves. 

In \cite{EGM00} we have demonstrated that 
the NNLO potential allows for a good description of the NN data, which 
is also visibly improved compared to the NLO results. Contrary to 
Ref.~\cite{Kaiser97},
we did not perform a  perturbative expansion of the {\it np} T--matrix 
in high partial waves and 
calculated phase shifts by solving the Lippmann--Schwinger (LS) equation for the 
NN T--matrix. We found that taking the momentum space cut--off 
in the LS equation of the order of 1 GeV 
allows for a satisfactory description of all partial waves simultaneously. 
With such a large value of the cut--off, the isoscalar central TPE potential  
becomes already so strongly attractive that unphysical deeply bound 
states appear in  the $D$-- and in the lower partial 
waves. Note that although such deeply bound states do not influence NN 
observables at low energies, they might show up in another processes (like e.g.~{\it Nd}
\cite{Epe02} and {\it $\pi$d} \cite{Beane02} scattering).
Since the potential is very strong (and attractive) and there are no counter 
terms according to the power counting, changing the value of the cut--off 
clearly leads to a strong variation of the $D$--wave phase shifts.
Higher order counter terms are needed in order to reduce the cut--off 
dependence of these observables and thus the problem with the slow 
convergence of the chiral expansion remains. 

Motivated by the known cancellation between the $\pi \pi$ and $\pi \rho$ 
exchanges, which has been observed in boson exchange models of the nuclear 
force, we  have constructed in Ref.~\cite{Epe02} the NNLO* version of the NN 
potential without spurious deeply bound states. To achieve that, we  adopted 
values of the LECs $c_{3,4}$, which are much smaller in magnitude 
than the ones obtained from $\pi$N scattering and result from 
subtracting the $\Delta$--contribution and fine tuning to NN observables:
$c_3 = -1.15$ GeV$^{-1}$,  $c_4 = 1.20$ GeV$^{-1}$. This also allowed for 
a fairly good description of the D-- and higher partial waves. Accounting 
for subleading TPE leads to small corrections in most channels and the  
chiral expansion for NN scattering seems to converge. 
Certainly, the situation is still far from being satisfactory since 
the small values of LECs $c_{3,4}$ are not compatible with  
$\pi$N scattering. Notice that the large values of these LECs are 
also supported by recent determinations from {\it pp} and {\it np} partial 
wave analysis performed by the Nijmegen group \cite{Rent99,Rent03}.

In the present work we explain the origin of the above mentioned problems and
present a way to improve the convergence
of the chiral expansion for the NN interaction. It allows to use the 
large values of $c_i$'s consistent with  $\pi$N scattering. 
We argue that the unphysically 
strong attraction in the isoscalar central part of chiral TPE at NNLO 
resulting when calculated using dimensional (or equivalent) regularization 
is due to high--momentum components of exchanged pions, which cannot be properly 
treated in an EFT. Using a cut--off regularization instead of dimensional one
and taking reasonable values for the momentum space cut--off allows to remove
spurious short--distance physics associated with high--momentum intermediate states 
and to greatly improve the convergence of the chiral expansion. A similar idea 
has already been used a long time ago in the analysis of the octet baryon masses and
the pion--nucleon sigma term \cite{JG} and was recently applied to improve the 
convergence of the SU(3) 
baryon chiral perturbation theory \cite{Don98,Don99,Bor02}. A critical
discussion about the use of cut-off schemes is provided in Ref.~\cite{BHM}. 
Notice further that a finite momentum--space cut--off in chiral loops has been used to 
derive the (energy--dependent) expressions for TPE in Ref.~\cite{Ordonez96}.
We also propose 
a simple and convenient way to derive analytic expressions for regularized 
TPE in the momentum space based on the spectral function representation.  

Our manuscript is organized as follows. In section \ref{sec2} we describe our 
formalism and present the explicit expressions for regularized TPE. 
In section \ref{sec3} we apply the formalism to {\it np} D-- and higher partial 
waves and compare the results with the ones obtained from dimensionally regularized 
expressions. The summary and conclusions are given in section \ref{sec4}.

\section{Formalism}
\def\theequation{\arabic{section}.\arabic{equation}}
\setcounter{equation}{0}
\label{sec2}

\begin{figure*}[htb]
\vspace{0.5cm}
\centerline{
\psfig{file=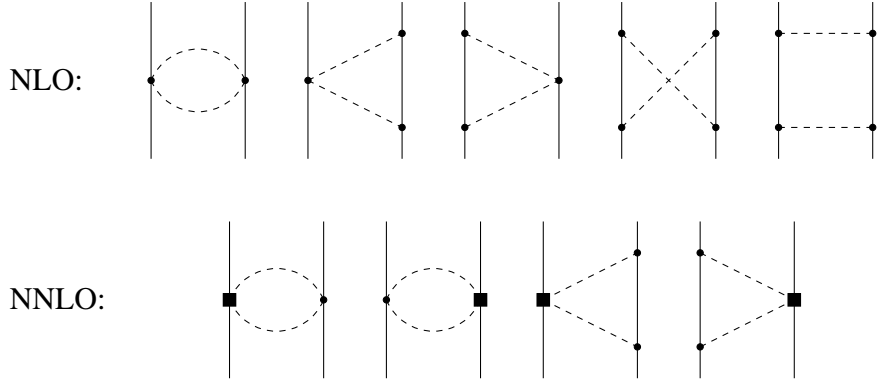,width=12cm}}
\vspace{0.3cm}
\centerline{\parbox{14cm}{
\caption[fig4]{\label{fig1} Chiral TPE at NLO and NNLO.
Heavy dots denote leading 
vertices from the chiral Lagrangian while solid rectangles correspond to 
subleading ones, which depend on the LECs $c_{1,3,4}$.  
}}}
\vspace{0.5cm}
\end{figure*}
Energy-independent expressions for the chiral TPE at NLO and NNLO have been 
derived using different formalisms in Refs.~\cite{Friar94,Kaiser97,EGM98}.
The corresponding diagrams are shown symbolically in Fig.~\ref{fig1}.
The last diagram in the first line (NLO) requires a special treatment in order 
to avoid  double counting of the iterated OPE. Notice further that the 
first two graphs in the second line lead to vanishing contributions to the 
NN force. The explicit time--ordered 
expressions for the potential can be found in \cite{EGM98}. While  OPE is of the order  
$(Q/\Lambda_\chi)^0$, where $\Lambda_\chi$ refers to the  chiral symmetry 
breaking scale of the order of the $\rho$--meson mass, TPE at NLO and NNLO
provides corrections of orders $(Q/\Lambda_\chi)^2$ and $(Q/\Lambda_\chi)^3$,
respectively. In the present work we will adopt the counting rule for the 
nucleon mass $m \gg \Lambda_\chi$, which is required for Weinberg's power counting
to be consistent \cite{Weinb1,Ordonez96}. Therefore, we do not need to include
relativistic $1/m$--corrections ($1/m^2$--corrections) to TPE (OPE) at the 
order considered. Notice that in the one--nucleon sector $m$ is 
usually treated on a same footing as $\Lambda_\chi$. The same counting rule has also 
been used in Ref.~\cite{Kaiser97}. The TPE contributions at NLO are given by:
\begin{eqnarray}
\label{2pi_nlo}
V^{\rm NLO} 
&=&  \frac{g_A^2}{(2 F_\pi)^4} \,  (\fet \tau_1 \cdot \fet \tau_2)  
\int \, \frac{d^3 l}{(2 \pi )^3}  \,
\frac{\left( {\vec l \,}^2 - {\vec q \, }^2 \right)}{\omega_+ 
\omega_- \left( \omega_+ + \omega_- \right)} 
\nonumber \\
&& {} - \frac{1}{8 (2 F_\pi)^4} \, ( \fet \tau_1 \cdot \fet \tau_2 ) \, \int  
\, \frac{d^3 l}{(2 \pi )^3}  \, 
\frac{ \left( \omega_+ - \omega_- \right)^2}{\omega_+ + \omega_-} 
\frac{1}{\omega_+ \omega_-} \nonumber   \\
&& {} - \frac{g_A^4}{2 (2 F_\pi)^4}  \,  
\int \, \frac{d^3 l}{ (2 \pi )^3} \,
\frac{\omega_+^2 + \omega_+ \omega_- + \omega_-^2}
{\omega_+^3 \omega_-^3 (\omega_+ + \omega_- )} \nonumber\\
&&\qquad \times \bigg\{  
 (\fet \tau_1 \cdot \fet \tau_2)\, 
\left( {\vec l \,}^2 
- {\vec q \,}^2 \right)^2 \nonumber \\
&& \qquad\quad + 6 ( \vec \sigma_2 \cdot [ \vec q \times \vec l \, ] ) 
( \vec \sigma_1 \cdot [ \vec q \times \vec l \, ] ) \bigg\} \; . 
\end{eqnarray}
where $\vec \sigma_i$ and  $\fet \tau_i$ are the spin-- and isospin--matrices of 
the nucleon $i$, $\vec q$ is the nucleon momentum transfer and 
$\omega_\pm = \sqrt{( \vec q \pm \vec l 
)^2 + 4 M_\pi^2 }$. The subleading TPE potential reads:
\beqa
\label{2pi_nnlo}
V^{\rm NNLO} 
&=&  \frac{3 g_A^2}{16 F_\pi^4} \;  \int    \frac{d^3 l}{(2 \pi )^3}  
\, \frac{\vec l \, ^2 - \vec q \, ^2}{\omega_-^2 \omega_+^2} \nonumber\\
&&\quad \,\times
\bigg( 8 c_1 \, M_\pi^2 + c_3 \, (\vec l \, ^2 - \vec q \, ^2) \bigg)
\, \nonumber\\
&& - \frac{c_4 \, g_A^2}{4  F_\pi^4}  
\, ( \fet \tau_1 \cdot \fet \tau_2 ) \, \int    \frac{d^3 l}{(2 \pi )^3}  
\, \frac{1}{\omega_+^2 \omega_-^2} \nonumber \\
&&\quad \,\times  ( \vec \sigma_2 \cdot [ \vec q \times \vec l \, ] ) 
 \times( \vec \sigma_1 \cdot [ \vec q \times \vec l \, ] )\,.
\eeqa
We will show in the next section how these integrals can be evaluated using 
different regularization schemes.

\subsection{Dimensional versus cut--off regularization}
\def\theequation{\arabic{section}.\arabic{equation}}

The integrals in Eqs.~(\ref{2pi_nlo}), (\ref{2pi_nnlo}) are ultraviolet divergent and 
thus need to be regularized. Applying dimensional regularization the TPE potential takes 
the form \cite{Kaiser97}:
\beq
\label{num1}
V_{\rm DR} =V_{\rm DR, \, non-pol.} + V_{\rm DR, \,  pol.} \, ,
\eeq
where the non--polynomial and polynomial (in $\vec q \,$) 
parts at NLO  $V_{\rm DR, \, non-pol.}^{\rm NLO}$ 
and  $V_{\rm DR, \,  pol.}^{\rm NLO}$, respectively,  read:
\beqa
\label{2pi_nlo_np}
V^{\rm NLO}_{\rm DR, \,  non-pol.} 
&=& - \frac{ \fet{\tau}_1 \cdot \fet{\tau}_2 }{384 \pi^2 F_\pi^4}\,
L(q) \, \biggl\{4M_\pi^2 (5g_A^4 - 4g_A^2 -1)\nonumber\\
&+& q^2(23g_A^4 - 10g_A^2 -1)
+ \frac{48 g_A^4 M_\pi^4}{4 M_\pi^2 + q^2} \biggr\}\nn
&-&  \frac{3 g_A^4}{64 \pi^2 F_\pi^4} \,L(q)  \, 
\Bigl[ ( \vec \sigma_1 \cdot \vec q \, ) \,
( \vec \sigma_2 \cdot \vec q \, ) 
\nonumber \\&& \quad - ( \vec \sigma_1 \cdot\vec
\sigma_2 ) \, q^2 \Bigr] ~,\\ 
V^{\rm NLO}_{\rm  DR, \,  pol.} 
&=& (S_1^{\rm DR} + S_2^{\rm DR} \,q^2) \, (\fet{
  \tau}_1 \cdot  \fet{ \tau}_2 ) \nonumber \\
&&+ S_3^{\rm DR} \, \Bigl[ ( \vec \sigma_1 \cdot \vec q \, ) \,
 ( \vec \sigma_2 \cdot \vec q \, ) - ( \vec \sigma_1 \cdot\vec
 \sigma_2 ) \, q^2 \Bigr]~. \nonumber
\eeqa
Here, we have set $q \equiv |\vec{q}\,|$ and the logarithmic loop function 
$L(q)$ is given by 
\beq\label{Lq}
L(q) = \frac{\omega}{q}\, 
\ln\frac{\omega + q}{2M_\pi}~,  \qquad \omega = \sqrt{q^2 + 4 M_\pi^2}\,.
\eeq
Further, 
\beqa 
\label{divs1}
S_1^{\rm DR}  &=& \frac{1}{384 \pi^2 F_\pi^4} \biggl\{ -18\Mpz (5g_A^4 -2g_A^2)
\ln \frac{\Mp}{\mu} + \alpha_1 \Mpz \biggr\}~,\nn
\label{divs2}
S_2^{\rm DR} &=& \frac{1}{384 \pi^2 F_\pi^4} \biggl\{ (-23g_A^4 +10g_A^2 +1)
\ln \frac{\Mp}{\mu} + \alpha_2  \biggr\},\nonumber \\
\label{divs3}
S_3^{\rm DR} &=& -\frac{3g_A^4}{64 \pi^2 F_\pi^4} \biggl\{ \ln
\frac{\Mp}{\mu} + \alpha_3\biggr\}~,
\eeqa
where $\mu$ is the scale of  dimensional regularization  and the $\alpha_i$ are 
polynomials in $g_A$, whose precise form depends on the choice of subtraction.
The non--polynomial part as well as all terms proportional to $\ln
({\Mp}/{\mu})$, which are  due to logarithmic divergences in
Eq.~(\ref{2pi_nlo}), are unique and do not depend on the choice of subtraction.
The $\mu$--dependence of the $V_{\rm DR, \, \rm  pol.}^{\rm NLO}$ is compensated
by the corresponding NLO counter terms of the form
\beqa
V^{\rm NLO}_{\rm cont} 
&=& C_1 (\mu) \, M_\pi^2 \,  (\fet{\tau}_1 \cdot  \fet{ \tau}_2 )
+ C_2 (\mu) \,q^2 \, (\fet{\tau}_1 \cdot  \fet{ \tau}_2 ) 
\nonumber \\
&+& C_3 (\mu)  \,  ( \vec \sigma_1 \cdot \vec q \, )  ( \vec \sigma_2 \cdot \vec q \, )
+C_4 (\mu) \, q^2 \,  ( \vec \sigma_1 \cdot\vec
 \sigma_2 ) ~,\nonumber \\ &&
\eeqa
and the resulting renormalized NN potential does not depend on $\mu$. 
Application of dimensional regularization to nuclear potentials is also 
discussed in \cite{Friar96}.

At NNLO one finds:
\beqa
\label{2pi_nnlo_np}
V^{\rm NNLO}_{\rm   DR, \, non-pol.} 
&=& -\frac{3g_A^2}{16\pi F_\pi^4}  \biggl\{2M_\pi^2(2c_1 -c_3) -c_3 q^2 \biggr\} 
\nonumber \\ && \quad \times (2M_\pi^2+q^2) A(q)  \nn
&& {} - \frac{g_A^2}{32\pi F_\pi^4} \,  c_4 (4M_\pi^2 + q^2) A(q)\, 
(\fet{ \tau}_1 \cdot \fet{ \tau}_2 ) \nonumber \\ && \times
\Bigl[ (\vec \sigma_1 \cdot \vec q\,)(\vec \sigma_2 \cdot \vec q\,) 
-q^2 (\vec \sigma_1 \cdot\vec \sigma_2 )\Bigr] \, \\ 
V^{\rm NNLO}_{\rm  DR, \,  pol.}  
&=& (\tilde S_1^{\rm DR}  + \tilde S_2^{\rm DR} \,q^2)  + \tilde S_3^{\rm DR}  \, (\fet{
  \tau}_1 \cdot  \fet{ \tau}_2 )\nonumber \\ 
&&\quad\times \Bigl[ ( \vec \sigma_1 \cdot \vec q \, ) \,
 ( \vec \sigma_2 \cdot \vec q \, ) - ( \vec \sigma_1 \cdot\vec
 \sigma_2 ) \, q^2 \Bigr]~, \nonumber
\eeqa
with the loop function $A(q)$
\beq\label{Aq}
A(q) = \frac{1}{2q} \arctan \frac{q}{2M_\pi}~,
\eeq
and
\beqa
\label{divs}
\tilde S_1^{\rm DR} &=& -\frac{3 g_A^2}{4 \pi F_\pi^4} (c_1 - c_3) M_\pi^3 \,, \nonumber\\
\tilde S_2^{\rm DR} &=& \frac{3 g_A^2}{16 \pi F_\pi^4} c_3 M_\pi\,, \\
\tilde S_3^{\rm DR} &=& -\frac{g_A^2}{32 \pi F_\pi^4} c_4 M_\pi\,.\nonumber
\eeqa
Notice that the integrals in Eq.~(\ref{2pi_nnlo}) are finite in dimensional 
regularization.

Although dimensional regularization provides an easy and convenient 
regularization scheme, it is by no means the only possible one. One can equally well 
regularize the divergent integrals in Eqs.~(\ref{2pi_nlo}), (\ref{2pi_nnlo})
using a momentum space cut--off, i.e.~by multiplying the 
corresponding integrands by the regulating function 
$f_\Lambda ( l ) \equiv f_\Lambda (| \vec l |)  $,
with the properties $f_\Lambda (l ) \stackrel{l \ll \Lambda}{\longrightarrow} 1$,
$f_\Lambda ( l ) \stackrel{l \gg \Lambda}{\longrightarrow} 0$. This function 
$f_\Lambda (l )$ should go to $0$ for large $l$ quick enough in order that
the regularized integrals exist. For cut--off regularized (CR) TPE one finds
similar to Eq.~(\ref{num1}):
\beq
\label{num2}
V_{\rm CR} = V_{\rm CR, \, non-pol.} + V_{\rm CR, \,  pol.}\,, 
\eeq
where 
\beqa
\label{condit}
V_{\rm CR, \, non-pol.}^{\rm NLO} &\stackrel{\Lambda \to \infty}{\longrightarrow}& \;
V_{\rm DR, \, non-pol.}^{\rm NLO} \,, \nonumber \\
V_{\rm CR, \, non-pol.}^{\rm NNLO} &\stackrel{\Lambda \to \infty}{\longrightarrow}& \;
V_{\rm DR, \, non-pol.}^{\rm NNLO} \,.
\eeqa
The polynomial pieces have the same structure 
as in Eqs.~(\ref{2pi_nlo_np}), (\ref{2pi_nnlo_np}),
where the quantities $S_i^{\rm DR}$ ($\tilde S_i^{\rm DR} $) should now be replaced by   
$S_i^{\rm CR}$ ($\tilde S_i^{\rm CR} $) given by 
\beqa 
\label{divs1c}
S_1^{\rm CR} &=& \frac{1}{384 \pi^2 F_\pi^4} \biggl\{ -18\Mpz (5g_A^4 -2g_A^2)
\ln \frac{\Mp}{\Lambda} \nonumber\\
&& \quad\qquad\quad + \alpha_1 ' \Mpz + \beta_1 \Lambda^2 \biggr\}~,\\
\label{divs2c}
S_2^{\rm CR} &=& \frac{1}{384 \pi^2 F_\pi^4} \biggl\{ (-23g_A^4 +10g_A^2 +1)
\ln \frac{\Mp}{\Lambda} + \alpha_2 '  \biggr\},\nonumber \\
\label{divs3c}
S_3^{\rm CR} &=& -\frac{3g_A^4}{64 \pi^2 F_\pi^4} \biggl\{ \ln
\frac{\Mp}{\Lambda} + \alpha_3 ' \biggr\}~,
\nonumber
\eeqa
and 
\beqa
\label{divsc}
\tilde S_1^{\rm CR} &=& -\frac{3 g_A^2}{4 \pi F_\pi^4} \Bigl( (c_1 - c_3) M_\pi^3 + 
\beta_2 M_\pi^2 \Lambda  + \beta_3 \Lambda^3 \Bigr) \,, \nonumber \\
\tilde S_2^{\rm CR} &=& \frac{3 g_A^2}{16 \pi F_\pi^4} \bigl(  c_3 M_\pi + \beta_4 
\Lambda  \bigr)\,, \\
\tilde S_3^{\rm CR} &=& -\frac{g_A^2}{32 \pi F_\pi^4} 
\bigl(  c_4 M_\pi + \beta_5 \Lambda  \bigr)\,.
\nonumber
\eeqa
Here $\alpha_i '$ are polynomials  in $g_A$ and $\beta_i$ are some combinations of 
$c_i$. The precise form of   $\alpha_i '$,  $\beta_i$ depends  on the choice of 
the regulating function $f_\Lambda (l)$. For a finite value of the cut--off  $\Lambda$
the function $V_{\rm CR, \, non-pol.}$ in Eq.~(\ref{num2}) contains, in general,  
not only non--polynomial terms in $q$ but also polynomial ones, 
which are however suppressed by inverse powers of the cut--off $\Lambda$. 
The terms in Eqs.~(\ref{divs1c}), (\ref{divsc}) proportional to $\Lambda$,  $\Lambda^2$ 
and $\Lambda^3$ correspond to linear, quadratic and cubic divergences in Eqs.~(\ref{2pi_nlo}),
(\ref{2pi_nnlo}) and are absent in the dimensionally regularized expressions.
Renormalization can be performed in very much the same way as before
by absorbing the terms  proportional to $\Lambda$,  $\Lambda^2$
and  $\ln \Lambda$ by the counter terms.  The only difference is that we 
now need the LO counter terms 
in order to get rid of the 
$\Lambda^2$--term in Eq.~(\ref{divs1c}) at NLO and the 
$\Lambda^3$--term in Eq.~(\ref{divsc}) at NNLO. 
In addition, NLO counter terms are required 
to renormalize the NNLO TPE potential. 
Notice that since  $V_{\rm CR, \, non-pol.}$
depends on $\Lambda$, the renormalized expressions 
for the potential using dimensional and cut--off regularizations are only identical 
with each other for $\Lambda \to \infty$.  Does that mean that we should 
necessarily take  $\Lambda \to \infty$? The answer is {\bf no}. This is because 
in an EFT one is usually only able to calculate observables with a finite accuracy 
performing calculations up to a certain order in the low--momentum expansion.
Taking $\Lambda \sim \Lambda_\chi$, the error from
keeping $\Lambda$ finite is beyond the theoretical accuracy. In other 
words, since 
\beq
V_{\rm CR, \, non-pol.} = V_{\rm DR, \, non-pol.} + \mathcal{O} (1/\Lambda) \,,
\eeq
the DR and CR expressions are identical up to higher order terms. At NNLO
one should choose $f_\Lambda (l)$ in such a way that 
$V_{\rm CR, \, non-pol.}^{\rm NLO} = V_{\rm DR, \, non-pol.}^{\rm NLO} 
+ \mathcal{O} (1/\Lambda^2)$.

\subsection{Spectral function representation}
\def\theequation{\arabic{section}.\arabic{equation}}

Before discussing implications of the choice of the regularization scheme on the 
convergence of the chiral expansion, we will show how cut--off regularization 
of the potential can be understood in terms of the spectral function representation. 
By their very nature, these spectral functions are the most natural objects 
to separate  the
long-- and short--distance contributions to the NN potential in momentum space.
We will switch to the notation 
introduced in Ref.\cite{Kaiser97} and express $V^{\rm NLO}$, 
$V^{\rm NNLO}$ as:
\beqa
V^{\rm NLO} &=& W_C (q) \, (\fet \tau_1 \cdot \fet \tau_2 ) + 
V_S (q) \, ( \vec \sigma_1 \cdot \vec \sigma_2 ) \nonumber \\ &+& V_T (q) 
\, ( \vec \sigma_1 \cdot \vec q \, ) \,( \vec \sigma_2 \cdot \vec q \, )\,, \nonumber\\
V^{\rm NNLO} &=& V_C (q) + 
W_S (q) \, (\fet \tau_1 \cdot \fet \tau_2 )\, ( \vec \sigma_1 \cdot \vec \sigma_2 ) 
\nonumber \\ &+& W_T (q) 
\,  (\fet \tau_1 \cdot \fet \tau_2 ) 
\, ( \vec \sigma_1 \cdot \vec q \, ) \,( \vec \sigma_2 \cdot \vec q \, )\,.
\eeqa
The functions $V_i (q)$ ($W_i (q)$) correspond to isoscalar (isovector) parts of the 
potential and can in case of dimensional regularization (and also for $\Lambda \to \infty$)
be read off from Eqs.~(\ref{2pi_nlo_np}), (\ref{2pi_nnlo_np}). 
The subscripts $C$, $S$ and $T$ stand for the central, spin--spin and 
tensor contributions, in order.

The functions  $V_i (q)$ ($W_i (q)$) can be represented (modulo terms polynomial in 
$q^2$)\footnote{In the present work we are only interested in the finite--range 
part of the two--pion exchange, which is given by the non--polynomial terms in 
momentum space.} 
by a continuous superposition 
of Yukawa functions \cite{Kaiser97,Chemt72}:
\beqa
\label{disp_int}
V_i (q) &=& \frac{2}{\pi} \int_{2 M_\pi}^\infty \, d \mu \, \mu \frac{\rho_{i} (\mu)}
{\mu^2 + q^2} \,, \nonumber\\
W_i (q) &=& \frac{2}{\pi} \int_{2 M_\pi}^\infty \, d \mu \, \mu \frac{\eta_{i} (\mu)}
{\mu^2 + q^2} \,,
\eeqa
where $\rho_i (\mu)$ and $\eta_i (\mu)$ are the corresponding mass 
spectra (spectral functions). Note that subtracted 
dispersion integrals should be used in  Eq.~(\ref{disp_int}) if  
spectral functions do not decrease for large $\mu$. This usually   
happens in the EFT calculations, where the spectral functions are obtained
within the low--$\mu$ expansion. Notice further that subtraction constants 
can be absorbed by the LECs corresponding to the short--range contact 
interactions and thus do not introduce any additional ambiguity. 
It is easy to see that 
$\rho_i (\mu)$ and $\eta_i (\mu)$ can be obtained from  $V_i (q) $ and  $W_i (q) $ via:
\beqa
\label{spectr:def}
\rho_i (\mu ) &=& {\rm Im} \Bigl[ V_i ( 0^+ - i \mu ) \Bigr]\,,\nonumber\\
\eta_i (\mu ) &=& {\rm Im} \Bigl[ W_i ( 0^+ - i \mu ) \Bigr]\,.
\eeqa
These spectral functions contain the whole dynamics corresponding to 
the exchanged $\pi\pi$ system. 
Once $\rho_i (\mu)$,  $\eta_i (\mu)$ functions are determined, the TPE 
potential can easily be obtained using Eq.~(\ref{disp_int}).

Let us now calculate the spectral function  $\rho_C^\Lambda (\mu)$, 
which results from the integral in  Eq.~(\ref{2pi_nnlo}) regularized
with a cut--off $\Lambda$. We will choose 
the regulating function $f_\Lambda (l)$ as 
$f_\Lambda (l) = \theta (\Lambda - l )$.
Performing integration over angles, one obtains
\beqa
V_C^\Lambda (q) &=&  \frac{3 g_A^2}{128 \pi^2  F_\pi^4} \; \int_0^\Lambda \, dl \, 
\frac{l (l^2 - q^2)}{q (l^2 + q^2 + 4 M_\pi^2)} \nonumber \\
&& \times 
\bigg( 8 c_1 \, M_\pi^2 + c_3 \, ( l^2 - q^2 ) \bigg)\\
&& \times
\biggl[\ln \Bigl(
(l+q)^2 + 4 M_\pi^2 \Bigr) 
 - \ln \Bigl((l-q)^2 + 4 M_\pi^2 \Bigr) \biggr]\,.\nonumber
\eeqa
One then finds for the spectral function
\beqa
\rho_C^\Lambda (\mu) &=& {\rm Im}  \Bigl[  V_C^\Lambda (  0^+ - i \mu )\Bigr] 
\nonumber \\
&=& -\frac{3 g_A^2}{64 F_\pi^4} \, 
\Big( 2M_\pi^2(2c_1 -c_3) + c_3 \mu^2 \Big)
(2 M_\pi^2 -\mu^2 )  \nonumber \\ && \quad 
\times  \frac{1}{\mu}\theta (\mu - 2 M_\pi)
\, \theta (\sqrt{\Lambda^2 + 4 M_\pi^2} - \mu )~.
\eeqa
The entire $\Lambda$--dependence is contained in the second Heaviside step--function,
which represents the only difference to the DR expression.
Thus, as one could expect for physical reasons, cutting off the momentum $l$ of 
one of the exchanged pions at $l = \Lambda$ leads to 
a cut--off in the TPE spectral functions, which in this specific  
case takes the value $\sqrt{\Lambda^2 + 4 M_\pi^2} \sim \Lambda$.
Clearly, the 
precise form of the resulting regulator in the spectral function representation depends 
on the choice of $f_\Lambda (l)$. 
Similar relations between the pion momentum and spectral function cut--offs can 
be obtained for other contributions to the NN potential as well. 
Using the regularized spectral function representation for TPE opens therefore 
an easy and convenient way to obtain the CR expressions and will 
be adopted in what follows. To be specific, we define the CR spectral functions
$\rho_i^\Lambda (\mu)$,  $\eta_i^\Lambda (\mu)$ according to 
\beqa
\label{reg_def}
\rho_i^\Lambda (\mu) = \rho_i (\mu) \; \theta (\Lambda - \mu) \,, \nonumber\\
\eta_i^\Lambda (\mu) = \eta_i (\mu) \; \theta (\Lambda - \mu)\,,
\eeqa
where $\rho_i (\mu)$,  $\eta_i (\mu)$ are the corresponding 
DR spectral functions, see section \ref{sec:CSR}.
The non--polynomial 
parts of the TPE potential at NLO and NNLO 
have then the same structure as in Eqs.~(\ref{2pi_nlo_np}),
(\ref{2pi_nnlo_np}), where the loop functions $L (q)$ and $A (q)$ should be replaced by 
$L^\Lambda (q)$ and $A^\Lambda (q)$ defined as:
\beqa
\label{def_LA}
L^\Lambda (q) &=& \theta (\Lambda - 2 M_\pi ) \, \frac{\omega}{2 q} \, 
\ln \frac{\Lambda^2 \omega^2 + q^2 s^2 + 2 \Lambda q 
\omega s}{4 M_\pi^2 ( \Lambda^2 + q^2)}~, \nonumber \\
s &=& \sqrt{\Lambda^2 - 4 M_\pi^2}\,,\nonumber \\
A^\Lambda (q) &=& \theta (\Lambda - 2 M_\pi ) \, \frac{1}{2 q} \, 
\arctan \frac{q ( \Lambda - 2 M_\pi )}{q^2 + 2 \Lambda M_\pi}\,.
\eeqa
Several comments are in order. 
First of all
we note that for $2 M_\pi < \Lambda$ the CR and DR expressions
$V_{\rm CR}$ and $V_{\rm DR}$ only differ from each other 
by higher order contact interactions (i.e. by short--range terms) 
if $V_{\rm CR}$ is expanded in powers of $1/\Lambda$. 
One can therefore use this regularization prescription in calculations at any given order  
in the low--momentum expansion without getting into trouble with spurious long--range 
contributions suppressed by inverse powers of $\Lambda$ which might arise for a different
choice of the cut--off function.
Further, one should keep in mind that our choice of  
regularization is by no means unique. Different choices lead to equivalent results for the 
potential (up to higher order terms) and may be used as well. 
Finally,  we would like to point out that the spectral function representation 
(\ref{disp_int}) does not allow to properly reproduce terms, which are polynomial 
in $q^2$ and non--analytic in $M_\pi^2$. 
Such terms are not important for our present work since we do not consider variation in 
$M_\pi$. If one is interested in the $M_\pi$--dependence of the nuclear force,
see for instance Ref.~\cite{EMG02}, a
cut--off regularization should be performed at the level of divergent integrals in 
Eqs.~(\ref{2pi_nlo}) and (\ref{2pi_nnlo}) rather than in the 
spectral function representation.

\subsection{Coordinate space representation}
\def\theequation{\arabic{section}.\arabic{equation}}
\label{sec:CSR}

The coordinate space representations $\tilde V_{C,S,T} (r)$  ($\tilde W_{C,S,T} (r)$)
of the isoscalar (isovector) central, spin--spin and tensor parts of the potential 
$V_{C,S,T} (q)$ ($W_{C,S,T} (q)$) are defined according to 
\beqa
\tilde V (r) &=& \tilde V_C (r) +  \tilde W_C (r) \; (\fet \tau_1 \cdot \fet \tau_2 ) 
\nonumber \\
&+& \Bigl(  \tilde V_S (r)   +  \tilde W_S (r) \; (\fet \tau_1 \cdot \fet \tau_2 ) \Bigr)
\, (\vec \sigma_1 \cdot \vec \sigma_2) \nonumber\\
&+& \Bigl(  \tilde V_T (r)   +  \tilde W_T (r) \; (\fet \tau_1 \cdot \fet \tau_2 ) \Bigr)
\nonumber \\ &&\quad \times (3 \vec \sigma_1 \cdot \hat r \; \vec \sigma_2 \cdot \hat r  - 
\vec \sigma_1 \cdot \vec \sigma_2 )\,.
\eeqa
The functions $\tilde V_{C,S,T} (r)$ 
can be obtained for any given $r > 0$ from the 
corresponding spectral functions via
\beqa
\label{fourc}
\tilde V_{C} (r) &=&  \frac{1}{2 \pi^2 r} \int_{2 M_\pi}^\infty d \mu \, \mu \, 
e^{-\mu r} \,\rho_{C} (\mu)\,, \\
\label{fourt}
\tilde V_{T} (r) &=& -\frac{1}{6 \pi^2 r^3}  \int_{2 M_\pi}^\infty d \mu \, \mu \, 
e^{-\mu r} \, ( 3 + 3 \mu r + \mu^2 r^2 ) \rho_T (\mu)\,, \nonumber \\ && \\
\label{fours}
\tilde V_{S} (r) &=&  -\frac{1}{6 \pi^2 r}  \int_{2 M_\pi}^\infty d \mu \, \mu \, 
e^{-\mu r} \, \Bigl( \mu^2 \rho_T (\mu) - 3 \rho_S (\mu ) \Bigr)\,. \nonumber \\ &&
\eeqa
The coordinate space representation of the isovector parts is given 
by the above equations replacing 
$\tilde V_{C,S,T} (r) \to \tilde W_{C,S,T} (r)$ and 
$\rho_{C,S,T} (\mu) \to \eta_{C,S,T} (\mu)$. 

We will now consider the coordinate space representation of the CR TPE potential.
According to the definition (\ref{reg_def}), Eqs.~(\ref{2pi_nlo_np}), 
(\ref{2pi_nnlo_np}) and (\ref{spectr:def}) we obtain
for the CR spectral functions at NLO
\beqa
\label{spectr_nlo}
\eta_{C}^{\Lambda}  (\mu) &=& 
 \frac{ 1}{768 \pi F_\pi^4}\,
 \, \biggl\{4M_\pi^2 (5g_A^4 - 4g_A^2 -1) 
\nonumber \\ && \quad - \mu^2(23g_A^4 - 10g_A^2 -1)
+ \frac{48 g_A^4 M_\pi^4}{4 M_\pi^2 - \mu^2} \biggr\} \, 
\nn
&& {} \times
\frac{\sqrt{\mu^2 - 4 M_\pi^2}}{\mu} \,
\theta ( \Lambda - \mu ) \\
\rho_{T}^{\Lambda} (\mu)  &=& \frac{1}{\mu^2}\, \rho_{S}^\Lambda  (\mu) 
=  \frac{3 g_A^4}{128 \pi F_\pi^4} \,
\frac{\sqrt{\mu^2 - 4 M_\pi^2}}{\mu} \,
\theta ( \Lambda - \mu )\,,
\nonumber
\eeqa
and  at NNLO 
\beqa
\label{spectr_nnlo}
\rho_{C}^{\Lambda}  (\mu) &=& -\frac{3 g_A^2}{64 F_\pi^4} \, \Big( 
2M_\pi^2(2c_1 -c_3) + c_3 \mu^2 \Big)\nonumber \\
&& \quad\times (2M_\pi^2 - \mu^2)
\, \frac{1}{\mu} \, \theta (\mu - 2 M_\pi)
\, \theta (\Lambda - \mu )\,,
\nonumber \\
\eta_{T}^{\Lambda}  (\mu) &=& \frac{1}{\mu^2}\, \eta_{S}^\Lambda  (\mu) =
- \frac{g_A^2}{128 F_\pi^4} \,  c_4 (4M_\pi^2 - \mu^2) \nonumber \\ &&\quad \times
\frac{1}{\mu} \, \theta (\mu - 2 M_\pi)
\, \theta (\Lambda - \mu )  
\,.
\eeqa

\begin{figure}[tb]
\centerline{
\psfig{file=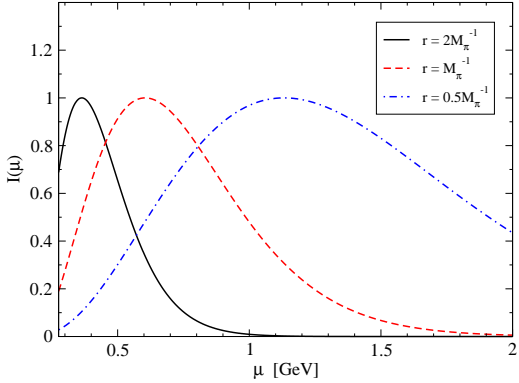,width=7cm}}
\vspace{0.3cm}
\centerline{\parbox{7cm}{
\caption[fig4]{\label{fig2}  The (normalized) integrand $I (\mu)$  in Eq.~(\ref{fourc}) for
different distances $r$. 
}}}
\end{figure}

We are now in the position to discuss implications of keeping the momentum space
cut--off $\Lambda$  in the above expressions for the spectral functions finite. Let us 
consider, for example, the isoscalar central part of the TPE. In Fig.~\ref{fig2} we 
show the (normalized) integrand $I (\mu)$ in Eq.~(\ref{fourc}) as a 
function of $\mu$ for 
$r= \frac{1}{2} M_\pi^{-1}$, $r= M_\pi^{-1}$, $r= 2 M_\pi^{-1}$ and for $\Lambda \to \infty$.
For the LECs $c_{1,3}$ we use the values \cite{Paul}: 
$c_1 =-0.81$ GeV$^{-1}$,  $c_3 =-4.7$ GeV$^{-1}$.
As expected, at large distances $r \geq 2 M_\pi^{-1}$ the integral in Eq.~(\ref{fourc}) 
is dominated
by low--$\mu$ components in the spectrum, where the chiral expansion for the spectral 
function is well behaved.
However, at intermediate distances $r \sim 1/M_\pi \sim 1.4$ fm the dominant 
contribution to the integral comes already from the region $\mu \sim 0.6$ GeV, where only a very 
slow  (if at all) convergence of the chiral expansion for $\rho (\mu)$ is expected. 
Certainly, at even shorter distances the resulting TPE potential is completely determined by 
the region of large $\mu$, where the spectral function is not properly described in chiral EFT. 
It is then clear that setting the upper limit of the integral in  Eq.~(\ref{fourc}) to infinity, 
which corresponds to the DR result, leads to inclusion of spurious short--range physics
in the TPE potential. We stress again that this problem solely arises because
at the order we are working there are no contact terms which normally would absorb these
contributions.
On the other hand,
introducing a finite cut--off $\Lambda$ in the spectral function representation according to 
Eq.~(\ref{reg_def}) we explicitly exclude all short--range components (i.e. those ones 
with the range $R < \Lambda^{-1}$) from the TPE potential, 
which are still present in the DR expression. The procedure is legitimate and does not lead
to any ambiguity if $\Lambda$ is chosen to be of the 
order of (or larger than) $\Lambda_\chi \sim M_\rho$, which provides a natural scale for the 
effective field theory. Various choices for $\Lambda$ lead to exactly the same 
result for low--energy observables provided one keeps terms in all orders in the EFT 
expansion. 
Choosing a specific value for $\Lambda$ in the calculation at any finite order in the low--momentum 
expansion one implicitly makes a particular choice for the combination of the higher order 
contact terms. There are obviously no restrictions in choosing $\Lambda$ if all LECs are of the 
natural size and the expansion parameter in EFT is small. 
Both dimensional and cut--off regularizations lead to similar results and 
the unphysical short--range components of TPE resulting from keeping 
$\Lambda$ very large (or even $\infty$) are compensated by corresponding counter 
terms to the order at which one is working. In fact, as we will show in the next section, 
the NN potential at NLO may serve as an example of such a situation. 

In certain cases, however, it appears to be advantageous to 
explicitly remove the spurious short--distance physics when calculating chiral loops. 
This happens, for instance, in SU(3) baryon chiral perturbation theory, where 
a much slower convergence of the chiral expansion is expected due to the relatively 
large mass of the strange quark. The leading nonanalytic components from loop corrections
calculated using dimensional regularization in some cases seem to destroy the good 
agreement of the lowest order calculation with data (like e.g. the Gell-Mann--Okubo relation
for the baryon masses).  Of course, this disagreement is corrected after 
inclusion of higher order contributions. However, the chiral expansion seems not to behave 
well and no clear convergence can be observed to the orders yet calculated
(such a statement holds e.g. for the baryon masses \cite{BoMe} but not for the 
baryon magnetic moments \cite{MeSt,KuMe}). Reformulating 
chiral EFT using a finite cut--off regularization allows to remove the spurious short--distance 
physics and to improve the convergence \cite{Don98,Don99,Bor02}, if the cut--off procedure
is implemented model--independently (as shown in Ref.~\cite{BHM}). 

As discussed in the introduction, slow convergence of the chiral series for the NN interaction
is observed if one uses the values for the LECs $c_{1,3,4}$ obtained in $\pi N$ scattering. 
The numerically large values of $c_{3,4}$ lead 
to a strong and attractive TPE at NNLO. Strong deviations from the NPSA results are observed in 
the $D$-- and $F$--waves which are parameter--free at this order and 
are still sensitive to the TPE contribution. We will demonstrate in the next section that the 
problems with the convergence are not due to the large values of $c_{3,4}$ which 
provide a proper long--range part of TPE \cite{Rent99,Rent03},
but rather due to unphysical short--range components in the DR expressions for the potential, 
which can (and should) be avoided using the CR.  

\begin{figure*}[htb]
\vspace{0.5cm}
\centerline{
\psfig{file=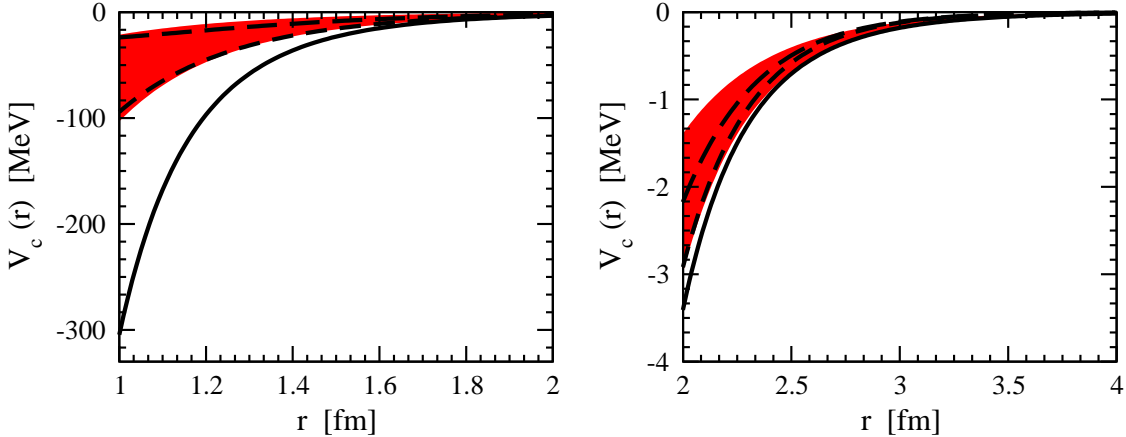,width=15.5cm}}
\vspace{0.3cm}
\centerline{\parbox{15.5cm}{
\caption[fig4]{\label{fig3}  The isoscalar central TPE potential at NNLO in $r$--space. 
The solid line shows the DR result corresponding to $\Lambda = \infty$ while the red band
results from varying $\Lambda$ between 500 and 800 MeV. The short-- (long--) dashed line 
shows the phenomenological $\sigma$ ($\sigma + \omega + \rho$) contributions based on the
isospin triplet configuration space version (OBEPR) of the Bonn potential \cite{Mach_rep}.   
}}}
\end{figure*}

To close this section 
we will give analytical expressions for TPE in the coordinate 
space. For the NNLO contributions one finds:
\beqa
\label{coord_nnlo}
\tilde V_C^\Lambda (r) &=&  \frac{3 g_A^2}{32 \pi^2 F_\pi^4} \, \frac{e^{- 2 x}}{r^6}
\bigg[ 2 c_1 \, x^2 (1 + x^2)^2 
\nonumber \\ && + c_3 (6 + 12x + 10 x^2 + 4 x^3 + x^4) \bigg]\nn
&& {} - \frac{3 g_A^2}{128 \pi^2 F_\pi^4} \, \frac{e^{- y}}{r^6} \bigg[
4 c_1 x^2 \Big(2 + y ( 2 + y) - 2 x^2   \Big) \nonumber \\
&& {} + c_3 \Big(  24 +  y (24 + 12 y  + 4 y^2 + y^3   )\nonumber \\
&& - 4 x^2 (2 + 2 y  + y^2 ) + 4  x^4 \Big) \bigg]\,, \\
\tilde W_T^\Lambda (r) &=& - \frac{g_A^2}{48 \pi^2 F_\pi^4} \, \frac{e^{- 2 x}}{r^6}
c_4 \, (1 + x) (3 + 3 x + x^2)  \nonumber\\
&& {}  + \frac{g_A^2}{768 \pi^2 F_\pi^4} \, \frac{e^{- y}}{r^6}
c_4 \, \Big( 48 + 48 y + 24 y^2 + 7 y^3 + y^4 
\nonumber \\ && \quad - 4 x^2 ( 8 + 5 y + y^2) \Big)\,,
\\
\tilde W_S^\Lambda (r) &=&  \frac{g_A^2}{48 \pi^2 F_\pi^4} \, \frac{e^{- 2 x}}{r^6}
c_4 \, (1 + x) (3 + 3 x + 2 x^2)\nonumber  \\
&& {}  - \frac{g_A^2}{384 \pi^2 F_\pi^4} \, \frac{e^{- y}}{r^6}
c_4 \, \Big( 24 + 24 y + 12 y^2 + 4 y^3 + y^4 
\nonumber \\ && \quad
- 4 x^2 ( 2 + 2 y + y^2) \Big)\, ,
\eeqa
with the abbreviations $x = M_\pi r$, $y = \Lambda r$.
For the NLO contributions we could not perform integrations in Eqs.~(\ref{fourc})--(\ref{fours})
analytically. In the chiral limit ($M_\pi = 0$) the results take, however, 
the following simple form:
\beqa
\label{coord_nlo}
\tilde W_C^\Lambda (r)\bigg|_{M_\pi=0} &=& \frac{23 g_A^4 - 10 g_A^2 - 1}{1536  \pi^3  F_\pi^4  r^5} \,
\bigg[ -6  \nonumber \\ && 
+ e^{-y} (6 + 6 y + 3 y^2 + y^3) \bigg] \,, \\ 
\tilde V_T^\Lambda (r)\bigg|_{M_\pi=0} &=& \frac{g_A^4}{256  \pi^3  F_\pi^4  r^5} \,
\bigg[ -15 \nonumber \\ &&   + e^{-y} (15 + 15 y + 6 y^2 + y^3) \bigg] \,, \\ 
\tilde V_S^\Lambda (r)\bigg|_{M_\pi=0} &=&  \frac{g_A^4}{128  \pi^3  F_\pi^4  r^5} \,
\bigg[ 6  \nonumber \\ &&  - e^{-y} (6 + 6 y + 3 y^2 + y^3) \bigg] \,. 
\eeqa
Certainly, the large--distance asymptotical behavior of the potential is unaffected by the cut--off 
procedure provided that $\Lambda \gg M_\pi$. 
In Fig.~\ref{fig3} we compare the isoscalar central part of TPE obtained using CR and DR.
The strongest effects of the cut--off are observed at intermediate ($\sim M_\pi^{-1}$)
and smaller distances, where TPE becomes unphysically attractive if DR is used. In contrast,
removing the large components in the mass spectrum of the TPE with the reasonably chosen cut--off 
$\Lambda =500 \ldots 800$ MeV greatly reduces this unphysical attraction and the resulting 
potential is of the same order in magnitude as the one obtained in phenomenological 
boson--exchange models. 
It remains to say that the CR expressions for the TPE are still not regular in the origin,
although the short--distance behavior is milder than in the case of DR and the leading 
singularities at $r =0$ are removed. For example, while $\tilde V_C (r) \propto 1/r^6$,
$\tilde V_C^\Lambda (r) \propto 1/r^5$ for $r \to 0$. The Schr\"odinger (or LS) equation 
has still to be regularized by introduction of an additional cut--off.

\section{Chiral TPE at NNLO and peripheral NN scattering}
\def\theequation{\arabic{section}.\arabic{equation}}
\setcounter{equation}{0}
\label{sec3}

In this section we will apply the OPE and TPE nuclear force at NNLO, calculated 
using the cut--off regularization as described in the previous section, to NN
phase shifts with orbital angular momentum $l \geq 2$ and to mixing angles 
with $j \geq 2$. 
As already pointed out in the introduction, no contact terms contribute at NNLO to the scattering 
amplitude in these channels. Consequently, such peripheral phase shifts are entirely
determined by the long--range part of the nuclear force and thus provide a sensitive test 
of the chiral TPE. The OPE potential at NLO (and NNLO) is given by 
\beq
V_{\rm OPE} = - \frac{g_A^2}{4 F_\pi^2} \bigg( 1 - \frac{4 M_\pi^2}{g_A} d_{18} \bigg)
\, \fet \tau_1 \cdot \fet \tau_2 \, \frac{(\vec \sigma_1 \cdot \vec q \,) 
(\vec \sigma_2 \cdot \vec q \,)}{q^2 + M_\pi^2}\,,
\eeq 
where the LEC $d_{18}$ is related to the Goldberger--Treiman discrepancy via
\beq
\frac{g_{\pi N}}{m} = \frac{g_A}{F_\pi} \bigg(1 - \frac{2 M_\pi^2}{g_A} d_{18} \bigg) \,.
\eeq
In what follows we use $g_A = 1.26$, $F_\pi =92.4$ MeV, $d_{18}=-0.97$ GeV$^{-2}$,
which leads to $g_{\pi N} \simeq 13.2$. For TPE at NLO and NNLO we use  
Eqs.~(\ref{2pi_nlo_np}), (\ref{2pi_nnlo_np}) with the functions $L (q)$, $A(q)$ replaced by 
$L^\Lambda (q)$, $A^\Lambda (q)$ defined in Eq.~(\ref{def_LA}). 
For the $c_i$'s we adopt the values \cite{Paul}:
$c_1=-0.81$ GeV$^{-1}$, $c_3=-4.70$ GeV$^{-1}$, $c_4=3.40$ GeV$^{-1}$.

The explicit expressions for the partial--wave decomposition of the potential can be 
found in Ref.~\cite{EGM00}. 
The partial--wave projected Lippmann--Schwinger (LS) equation for the NN T--matrix reads:
\beqa
\label{LSeq}
T_{l,\, l'}^{s  j} ( p \, ', \,  p) &=&
V_{l,\, l'}^{s  j} ( p \, ', \,  p) + \sum_{l''} \int \,
\frac{d^3 p''}{(2 \pi)^3} V_{l,\, l''}^{s  j} ( p \, ', \,  p'')
\nonumber \\ &&\times \frac{m}{ p\, ^2 -  (p '')^2 + i \epsilon} 
T_{l'',\, l'}^{s  j} ( p'' , \,  p)\,,
\eeqa
where the on--shell S-- and T--matrices are related via
\beq
S_{l, \, l'}^{s  j} (p,  \,  p) = \delta_{l  \, l'} \, 1^{sj}
- \frac{i}{8 \pi^2}\, p\, m \, T_{l,\, l'}^{s  j} 
( p , \,  p )\,.
\eeq
Since the  chiral potential grows at large momenta, 
the LS equation (\ref{LSeq}) has to be regularized,\footnote{This 
regularization of the LS equation should not be confused with the CR of the chiral loops
discussed in previous sections.} which requires introduction of an additional cut--off
(which can be done consistently with the CR discussed so far). 
It is commonly believed (and observed in various boson--exchange and phenomenological potential 
models, see also Ref.~\cite{Epe02}) 
that because of the centrifugal barrier the NN interaction in the peripheral partial waves
becomes weak enough to be treated perturbatively. 
This is also confirmed by the smallness of the corresponding
phase shifts. To calculate phase shifts in high partial waves one may therefore 
use the Born approximation to the T--matrix:\footnote{Notice, however, that the weakness 
of the NN interaction for high values 
of $l$ is not related to the chiral expansion and does not follow from the power counting.}
\beq
T_{l,\, l'}^{s  j} ( p \, ', \,  p) \sim
V_{l,\, l'}^{s  j} ( p \, ', \,  p)\,.
\eeq
Such a procedure, which is analogous to the one of Refs.~\cite{Kaiser97,Ent02}, 
allows to avoid the introduction of an additional cut--off in the LS equation
and will be adopted in the present work. One should, however, always keep in mind that
this approximation breaks down if the phase shifts become large. 

\begin{figure*}[htb]
\vspace{0.5cm}
\centerline{
\psfig{file=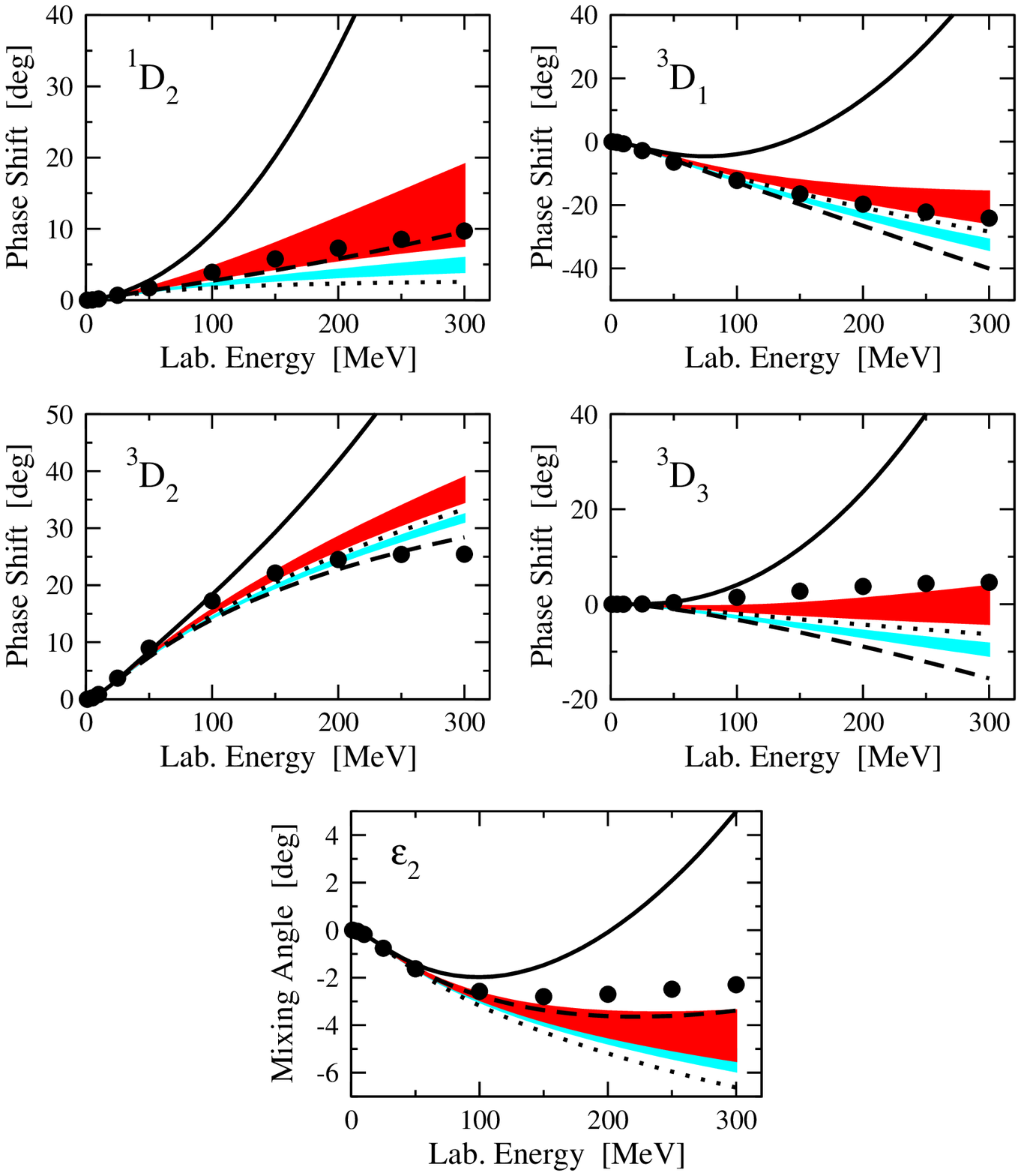,width=15.cm}}
\centerline{\parbox{15.5cm}{
\caption[fig4]{\label{fig4} D--wave NN phase shifts and mixing angle $\epsilon_2$ versus
the nucleon laboratory energy. The dotted curve is the LO result (i.e. pure OPE), while the dashed
(solid) curve refers to NLO (NNLO) results for OPE+TPE with the potential calculated using 
dimensional regularization. The light (dark) shaded band shows the NLO (NNLO) predictions with 
chiral TPE obtained using the cut--off regularization with $\Lambda =500 \ldots 800$ MeV.
The filled circles depict the Nijmegen PSA results \cite{NPSA}.
}}}
\end{figure*}

Before presenting our predictions for the high partial waves, we would like to specify the 
differences between our analysis and the one of Kaiser et al. \cite{Kaiser97}:
\begin{itemize}
\item
First of all, we strictly follow the lines of Weinberg's power counting and do not 
include relativistic corrections to the nuclear force at NNLO, which have been included 
in Ref.~\cite{Kaiser97}. In contrast to this reference, we use non--relativistic kinematics 
when calculating phase shifts.
\item
We also do not include the contribution from once iterated OPE, which turns out to be 
numerically small in most channels. From the point of view of the power counting, 
there is no reason to include once iterated OPE and not to include two--, three--, $\ldots$
times iterated OPE. 
\item
We use slightly different (and more modern) values for the LECs $c_{1,3,4}$ and for $g_A$. 
\item
The last and most important difference is that we use the finite cut--off $\Lambda$ to regularize
chiral loop integrals. This is in strong contrast with the analysis of Ref.~\cite{Kaiser97},
where DR corresponding to $\Lambda= \infty$ has been adopted. As it is clear from the 
above discussion, optimal values for the cut--off are those close to the scale where 
the EFT description becomes inaccurate. Taking a too small $\Lambda$ will remove the truly 
long--distance physics while too large values for the cut--off may affect the convergence 
of the EFT expansion due to inclusion of spurious short--distance physics. 
We will therefore vary $\Lambda$ in the range $\Lambda= 500 \ldots 800$ MeV which appears 
to be physically reasonable and matches well with both $M_\rho$ and the cut--off used in the 
LS equation for NN scattering \cite{Epe02}. 
\end{itemize}


\begin{figure*}[tb]
\vspace{0.5cm}
\centerline{
\psfig{file=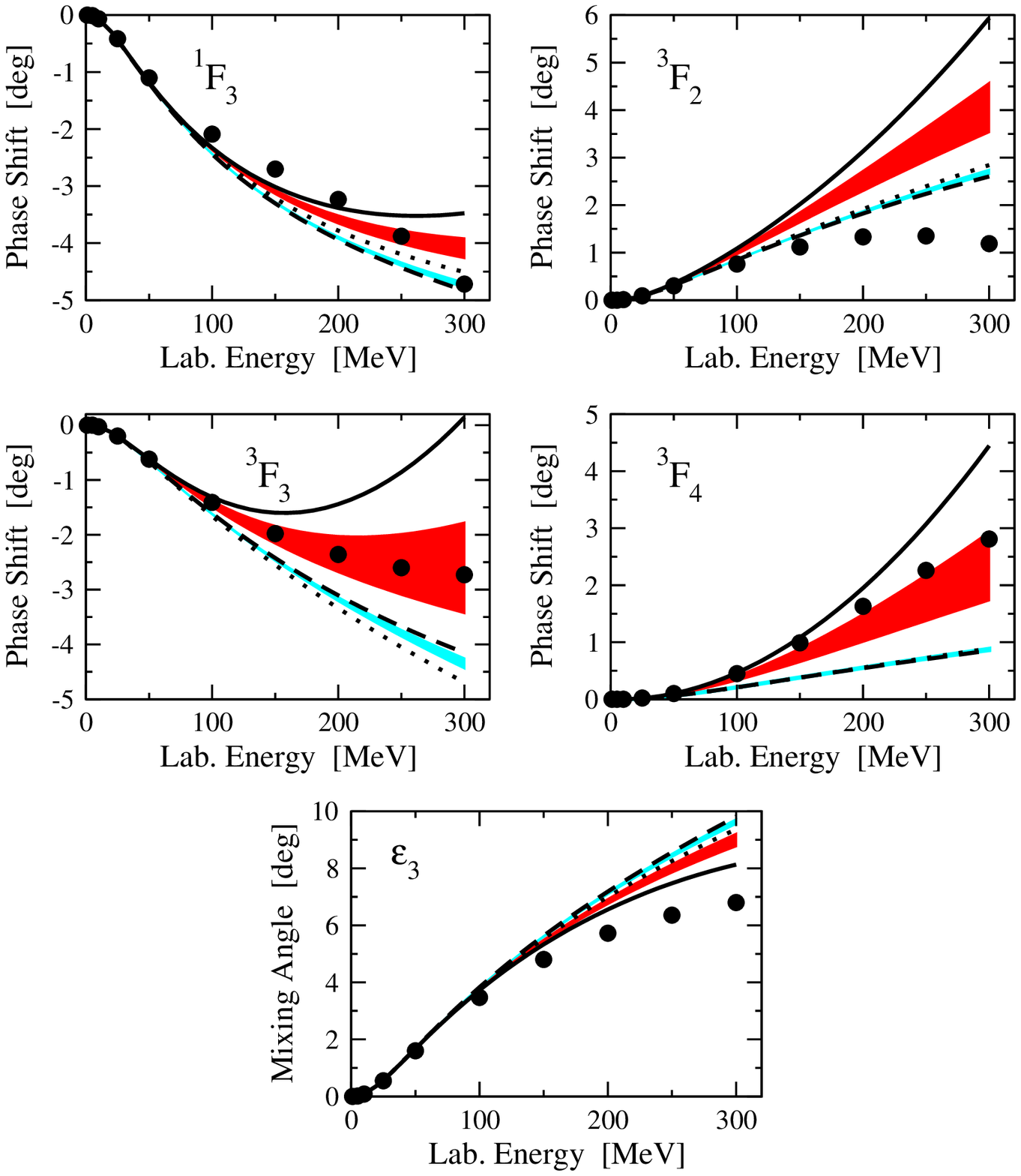,width=15.cm}}
\centerline{\parbox{15.5cm}{
\caption[fig5]{\label{fig5}  F--wave NN phase shifts and mixing angle $\epsilon_3$ versus
the nucleon laboratory energy. For notation see  Fig.~\ref{fig4}.
}}}
\end{figure*}

Let us start with the D--waves which are shown in Fig.~\ref{fig4}. The LO result represented by 
pure OPE already provides a good approximation to the phase shifts in the $^3D_1$ and $^3D_2$ partial
waves and to the mixing angle $\epsilon_2$. It is too weak in the $^1D_2$ channel and does not describe
properly the $^3D_3$ phase shift.  The letter appears to be quite small ($| \delta | \sim 4.6 ^\circ$ at 
$E_{\rm lab} = 300$ MeV) compared to the other D--wave phase shifts ($| \delta | \sim 9.7 ^\circ
\ldots 25.5 ^\circ$). The reason is that partial--wave projected OPE, taken on the energy shell, 
is strongly suppressed in this channel. Consequently, the  $^3D_3$ phase shift is quite sensitive
to TPE but also to the iteration of the potential which we neglect in the present analysis.
\begin{figure*}[tb]
\vspace{0.5cm}
\centerline{
\psfig{file=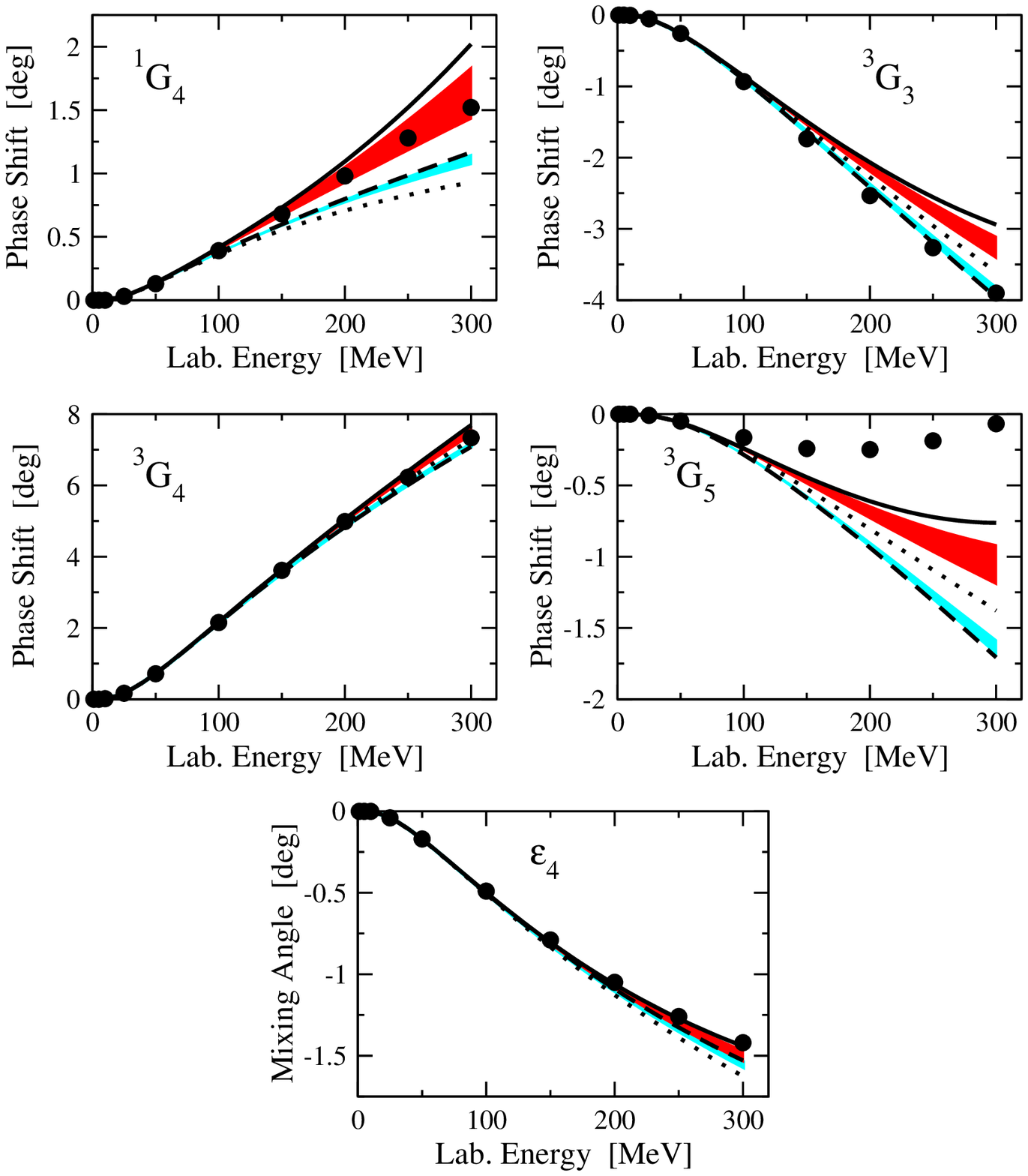,width=15.cm}}
\centerline{\parbox{15.5cm}{
\caption[fig6]{\label{fig6}  G--wave NN phase shifts and mixing angle $\epsilon_4$ versus
the nucleon laboratory energy. For notation see  Fig.~\ref{fig4}.
}}}
\end{figure*}

The NLO predictions obtained using dimensional regularization are shown by the dashed curves.
One observes a visible improvement for the $^1D_2$ phase shift and for $\epsilon_2$ while the 
NLO corrections go 
in the wrong direction in the $^3D_1$ and  $^3D_3$ channels. The NNLO predictions calculated 
with  dimensional regularization are depicted by the solid lines. The good agreement with the 
data observed at LO and NLO is destroyed in all partial waves for energies $E_{\rm lab} > 50$ 
MeV and the chiral expansion does not seem to converge. Note that at N$^3$LO one independent 
contact operator contributes to each D--wave so that the agreement with the data will presumably
be restored, see also Ref.~\cite{EGM00} for a related discussion.
Note that the results presented here are parameter--free and are very similar to the ones of 
Ref.~\cite{Kaiser97}.
No breakdown of the chiral expansion at NNLO is observed using the CR in the chiral loops and 
choosing $\Lambda = 500 \ldots 800$ MeV. This proves explicitly 
that strong disagreement with the data resulting from chiral TPE at NNLO 
is due to unphysical short--distance components, which are kept in the 
DR expressions for the potential. 
The use of a momentum space cut--off keeps only the long--distance portion
of the chiral loops and leads to a greatly improved behavior.
Our predictions agree with the data in the  $^1D_2$ 
and $^3D_1$ channels and go into the right direction in the $^3D_3$ partial wave and for 
$\epsilon_2$. Note that the uncertainty due to the variation of $\Lambda$ is quite significant 
at higher energies. The NLO bands are narrower than the NNLO ones as a consequence of the fact
that the leading TPE contribution is numerically quite small.

Our predictions for the F--wave phase shifts and for $\epsilon_3$ are presented in Fig.~\ref{fig5}. 
Although the situation with  DR subleading TPE is not as dramatic as for D--waves, a too strong
attraction is clearly visible in the $^3F_2$, $^3F_3$ and $^3F_4$ partial waves. Removing the 
short--distance components of TPE with the cut--off regularization leads to a significant 
improvement in the  $^3F_3$ and $^3F_4$ channels, while additional repulsion is still missing in 
the $^3F_2$ partial wave.

For the G-- and higher partial waves we observe very similar results for both DR and CR TPE 
and the phase shifts are essentially given by OPE. Notice that additional attraction is  
generated from iterated OPE in the $^3G_5$ partial wave \cite{Kaiser97}, 
which is not included in the present work. Therefore, the strong disagreement with the data in 
this channel is presumably just an artefact of the Born--approximation.

Finally, we would like to point out that the typical uncertainty of $\sim 30$\% observed in 
our NNLO predictions at $E_{\rm lab} =300$ MeV is consistent with the power counting. 
Indeed, the N$^3$LO counter terms are expected to provide corrections to the S--matrix 
of the order ${Q^4}/{\Lambda_\chi^4}$.
Taking $Q \sim 375$ MeV, which corresponds to $E_{\rm lab} =300$ MeV, and identifying 
$\Lambda_\chi$ with the smallest value used for the cut--off $\Lambda$, i.e. $\Lambda=500$ MeV,
one estimates the  N$^3$LO effects as $\sim 32$\%. As already stated before, the uncertainty 
is larger in the cases, where phase shifts are numerically small (like, for instance, in the 
$^1D_2$, $^3D_3$, $^3G_5$ partial waves). Although the following remark 
is quite obvious, we stress
that a two-nucleon potential based on a systematic EFT approach should not be fine-tuned to
fulfill a $\chi^2/{\rm dof } \simeq 1$ as it is done in more conventional approaches. This
does, however, not mean that such a precision can not be reached.

\section{Summary and conclusions}
\def\theequation{\arabic{section}.\arabic{equation}}
\setcounter{equation}{0}
\label{sec4}

In this paper we have considered the two-nucleon potential in chiral effective field theory,
making use of a novel method of regularizing the pion loop integrals. For that, we have
considered the spectral functions obtained from the NLO and NNLO TPE contributions and argued
that only masses below the chiral symmetry breaking scale should contribute in the loop integrals.
This can be easily implemented by applying a cut--off to the spectral functions.
Varying the
 cut--off between 500 and 800 MeV (as given by the mass of the heaviest Goldstone
boson, the eta, and the mass of the lightest resonance, the rho), 
we find the following pertinent results:

\begin{itemize}
\item[1)]From the regularized spectral functions, we have constructed the coordinate
space representations of the various components of the NN potential. The isoscalar
central TPE shown in Fig.~\ref{fig3} agrees with phenomenological potentials. The
strong unphysical attraction of the TPE is a short--range phenomenon which is suppressed
in CR by choosing $\Lambda \leq M_\rho$, whereas in the Bonn potential 
$\pi \rho$ exchange cancels the strong attractive TPE, which is one particular
model for this kind of short-range physics.
\item[2)]We have considered the peripheral partial waves ($l \ge 2$), 
because at NNLO, these are given entirely by OPE and TPE with no free parameters.
We have calculated these phases in Born approximation which should be legitimate
at least for the D-- and higher waves, see Ref.~\cite{Epe02} for more discussion. 
The uncertainty in most D-- and F--waves at NNLO is still sizeable 
even with the finite cut--off. 
This has to be expected because of the large values of $c_{i}$'s. The results for D-- 
and F--waves are still not completely converged at NNLO, but the error of about 
10~(1)$^\circ$ at $E_{\rm lab} = 300$ MeV for the D-- (F--)waves appears reasonable.  
There is no breakdown of the chiral expansion for D--waves beyond $T_{\rm lab} = 50$ MeV and 
for F--waves beyond   $T_{\rm lab} = 150$ MeV as found in Ref.~\cite{Kaiser97} 
using dimensional regularization.
\item[3)]It is no surprise that the  NNLO TPE gives larger corrections than the NLO one 
because of the delta contributions subsumed in the LECs $c_3$ and $c_4$. 
One might therefore contemplate  including the $\Delta (1232)$ explicitly in the
effective Lagrangian  (see e.g. \cite{Ordonez96}), because in such a theory most 
(but not all!) of the NNLO effects are shifted to NLO, provided that a systematic analysis
of pion-nucleon scattering in such a scheme is available (for attempts see 
e.g. \cite{ET1,ET2,FMdel}). However, for obtaining a precise potential one still would 
have to go to NNLO, which is considerably more complicated than in the pion--nucleon EFT, 
as witnessed by the fact that no complete fourth order calculation with deltas in the
single nucleon sector exists.
\item[4)] We stress that dimensional regularization is by no means ruled out by such 
considerations. In general, for quickly converging expansions, it should be the 
method of choice. If, however, the convergence for some well understood physical
reason is slow 
and (some) observables become sensitive to spurious short--distance physics kept in DR,
it might be preferable to use CR, as done here. 
In our case the LECs $c_{3,4}$ are large. Choosing DR, one generates 
a series of higher order contact interactions, which are  $\propto c_{3,4}$ 
and therefore large.  The  D-- and F--waves are certainly most sensitive to such contact 
interactions and are 
thus strongly affected. 
\item[5)] The cut--off used here is not a form--factor and 
is not related to
the finite extension of the nucleon. Obviously, the precise shape of the regulating function 
and the precise value of $\Lambda$ are not important (as long as the value for $\Lambda$
is 
well above the two--pion threshold, $\Lambda > 2M_\pi$ 
and below the scale of chiral symmetry breaking, $\Lambda < \Lambda_\chi$ ).
\end{itemize}

In a subsequent publication,  we will apply CR to the low partial waves in the 
non--perturbative regime,  where we have to solve the regularized Lippmann--Schwinger equation
Eq.~(\ref{LSeq}) to generate the bound and scattering states. 
It can be demonstrated  that there are no deeply bound states 
and low--energy observables are not affected by CR.

\subsection*{Acknowledgments}
\def\theequation{\arabic{section}.\arabic{equation}}

We would like to thank John Donoghue for important comments and Norbert Kaiser 
and Jambul Gegelia for useful discussions. This work is supported in part by the 
Deutsche Forschungsgemeinschaft (E.E.).


\begin{thebibliography}{99}
\bibitem{Ulf}U.-G.~Mei{\ss}ner,
Rept.\ Prog.\ Phys.\   56 (1993) 903.\vs
\bibitem{Pich}
A.~Pich,
Rept.\ Prog.\ Phys.\   58 (1995) 563.\vs
\bibitem{BKMrev}
V.~Bernard, N.~Kaiser and U.-G.~Mei{\ss}ner,
Int.\ J.\ Mod.\ Phys.\ E  4 (1995) 193.\vs
\bibitem{Ecker}
G.~Ecker,
Prog.\ Part.\ Nucl.\ Phys.\  36 (1996) 71.\vs
\bibitem{Weinb1} S. Weinberg, Phys. Lett. B 251 (1990) 288.\vs
\bibitem{Weinb2} S. Weinberg, Nucl. Phys.  B 363 (1991) 3.\vs
\bibitem{Ordonez96}C.~Ord\'{o}\~{n}ez, L.~Ray and U.~van Kolck,  Phys. Rev. 
        C 53 (1996) 2086.\vs
\bibitem{Friar94} J.L. Friar, S.A. Coon, Phys. Rev. C 49 (1994) 1272.\vs
\bibitem{Kaiser97} N. Kaiser, R. Brockmann, and W. Weise, Nucl. Phys. 
        A 625 (1997) 758. \vs
\bibitem{EGM98} E. Epelbaoum, W. Gl\"ockle, and U.-G. Mei\3ner, Nucl. 
        Phys. A 637 (1998) 107.\vs
\bibitem{BKM95} V. Bernard, N. Kaiser, and U.-G. Mei\3ner, Nucl. Phys. B 457 (1995) 147.\vs
\bibitem{BKM97} V. Bernard, N. Kaiser, and U.-G. Mei\3ner, Nucl  Phys. A 615 (1997) 483.\vs
\bibitem{Moi98} M. Moj\v{z}i\v{s}, Eur. Phys. J. {C2} (1998) 181.\vs
\bibitem{FMS99} N. Fettes, U.-G. Mei\3ner, and S. Steininger, Nucl. Phys. A 640 (1998) 199.\vs
\bibitem{Paul} P. B\"uttiker, and U.-G. Mei\3ner, Nucl. Phys. A 668 (2000) 97.\vs
\bibitem{FM4}
N.~Fettes and U.-G.~Mei{\ss}ner,
Nucl.\ Phys.\ A 676 (2000) 311.\vs
\bibitem{FM3}
N.~Fettes and U.-G.~Mei{\ss}ner,
Nucl.\ Phys.\ A  693 (2001) 693.\vs
\bibitem{Epe02} E.~Epelbaum, A.~Nogga, W.~Gl\"ockle, H.~Kamada, 
U.-G.~Mei{\ss}ner and H.~Witala,
Eur. Phys. J. A 15 (2002) 543.\vs
\bibitem{GGIP}A.~D.~Galanin, A.~F.~Grashin, B.~L.~Ioffe and I.~Ya.~Pomeranchuk,
Nucl.\ Phys.\ 17 (1960) 181.\vs 
\bibitem{Stoks93}  V.G.J. Stoks, R.A.M. Klomp, M.C.M. Rentmeester, and J.J. de Swart, 
Phys. Rev. C 48 (1993) 792.\vs
\bibitem{EGM00}  E. Epelbaum, W. Gl\"ockle, and U.-G. Mei\3ner, Nucl. Phys. A 671 (2000) 295.\vs
\bibitem{Beane02} S.R.~Beane, V.~Bernard, E.~Epelbaum, U.-G.~Mei{\ss}ner 
and D.~R.~Phillips, accepted for publication in Nucl. Phys. A.\vs
\bibitem{Rent99} M.C.M. Rentmeester, R.G.E. Timmermans, J.L. Friar, and J.J. de Swart, 
Phys. Rev. Lett. 82 (1999) 4992.\vs
\bibitem{Rent03} M.C.M. Rentmeester, R.G.E. Timmermans,  and J.J. de Swart, 
accepted for publication in Phys. Rev. C.\vs
\bibitem{JG} 
J.~Gasser,
Annals Phys.\  136 (1981) 62.\vs
\bibitem{Don98} J.F. Donoghue, and B.R. Holstein, Phys. Lett. B 436 (1998) 331.\vs 
\bibitem{Don99} J.F. Donoghue, B.R. Holstein, and B. Borasoy, Phys. Rev. D 59 (1999) 036002.\vs
\bibitem{Bor02} B. Borasoy, B.R. Holstein, R. Lewis, and P.-P.A. Ouimet, Phys. Rev. D 66 (2002) 094020.\vs
\bibitem{BHM}V. Bernard, T.R. Hemmert and U.-G. Mei{\ss}ner, ``Cutoff schemes in chiral
perturbation theory'', in preparation.\vs
\bibitem{Friar96} J.L. Friar, Mod. Phys. Lett. A11 (1996) 3043.\vs
\bibitem{Chemt72} M. Chemtob, J.W. Durso, and D.O. Riska, Nucl. Phys. B 38 (1972) 141.\vs
\bibitem{EMG02} E. Epelbaum, U.-G. Mei{\ss}ner, W. Gl\"ockle, Nucl.Phys. A714 (2003) 535. \vs
\bibitem{BoMe}
B.~Borasoy and U.-G.~Mei{\ss}ner,
Annals Phys.\   254 (1997) 192.\vs
\bibitem{MeSt}
U.-G.~Mei{\ss}ner and S.~Steininger,
Nucl.\ Phys.\ B  499 (1997) 349.\vs
\bibitem{KuMe}
B.~Kubis and U.-G.~Mei{\ss}ner,
Eur.\ Phys.\ J.\ C 18 (2001) 747.\vs
\bibitem{Mach_rep} R. Machleidt, K. Holinde and Ch. Elster, Phys. Rep. 149 (1987) 1.\vs
\bibitem{NPSA} V.G.J. Stoks, R.A.M. Klomp, M.C.M. Rentmeester, and  J.J. de Swart, 
Phys. Rev. C 48 (1993) 792.\vs
\bibitem{Ent02}  D.R. Entem, and R. Machleidt,  Phys. Rev. C 66 (2002) 014002.\vs
\bibitem{ET1}
P.~J.~Ellis and H.~B.~Tang,
Phys.\ Rev.\ C  56 (1997) 3363.\vs
\bibitem{ET2}
P.~J.~Ellis and H.~B.~Tang,
Phys.\ Rev.\ C  57 (1998) 3356.\vs
\bibitem{FMdel}
N.~Fettes and U.-G.~Mei{\ss}ner,
Nucl.\ Phys.\ A 679 (2001) 629.\vs
\end{thebibliography}
\end{document}